# Geometrical Field Representation of Solid, Fluid, and Gas as Continuum in Rational Mechanics

XIAO Jianhua

Henan Polytechnic University, Measurement Institute, Jiaozuo, Henan, 454000, China

**Abstract:** Any materials have three physical states as solid, liquid, and gas. Basically, the materials at some molecular cluster scales are assumed as the same for all the above physical states. So, a basic gauge tensor field can be attached to the basic molecular cluster elements (basic material element). Under this objective (material element) invariance, the concept of continuum in rational mechanics is studied in this paper. Different material phases have different basic gauge fields. So, the matter phases are expressed by the basic gauge transformations. Based on this general understanding, the different phases have different motion transformations, internally. Based on the points-set transformation concept about the motion transformation in continuum, the macro classical strain is expressed by the additive addition of the intrinsic stretching of material element and its intrinsic local rotation. For zero classical strain (no macro deformation observed on its configuration surface, suitable container is required for liquid and gas to make up macro invariant configuration), the results show that: (1) For solid, the local rotation angular is zero. The material element has no intrinsic stretching. (2) For liquid, the local rotation will not change the basic gauge tensor. The material element has intrinsic plane stretching on the rotation plane. (3) For gas state, the intrinsic local rotation will amplify the basic gauge tensor. The material element has intrinsic stretching along the rotation direction. Hence, under the condition of no macro classical strain be observed, the material element has three different physical states: solid (no intrinsic stretching), fluid (plane intrinsic stretching), and gas (directional intrinsic stretching). Furthermore, for the three states, the free conditions are defined by zero intrinsic stretching. Referring to this free condition, the constitutive equations for the materials at multiple states are established. It is shown that the classical constitutive equations are included in this general unified formulation.

**Contents:**

## 1. Introduction

In classical physics, the materials as a continuum are classified as solid, liquid, and gas. For each state, some corresponding mechanics theory or interpretations are formulated. The three states are not unified as a continuum in rational sense but it is in the conceptual sense. In this research, a rational unifying is persuaded. In fact, at actual world, many materials are composed of the combination of solid, liquid, and gas states. The minor gas or liquid states material in a working part of machine may cause the failure of the working part. So, for advanced machinery industry, a constitutive equation which taken the three states into consideration are urgently needed.

On physical view-point, although the three states are unified as continuum, the rational formulation which can be used to explain the intrinsic difference among them is still missing. Phenomenally, for idea solid state, the deformation of material configuration is assumed as zero without external action. So, it has no configuration change at natural state. Its deformation stress is completely by the classical strain. For liquid, to make it has a fixed configuration (externally observed), a suitable container must be used. The internal motion of material elements is not zero. So, a static compressive pressure parameter is introduced to express such a kind of internal motion. When the internal motion tends to be zero, it becomes to the solid state. So, the boundary between the solid phase and the liquid phase is not so clear as we are taught by the standard textbook.

If the internal motion is strong enough, a closed container must be used to keep the macro external configuration of gas. So, the gas has a natural expansion pressure. This is essentially different from the nature of liquid phase. However, the boundary between both is still not clear in rational sense.

By physical facts, two or three phases can exist spontaneously for a suitable temperature. As the temperature is the measure the internal motion of materials, the only conclusion which one can make out is that the internal motion is continuous although the different phase states seems to be so different in phenomenon sense. Therefore, a continuous motion representation should be used to unify the different phase states. Only in such a rational formulation, the continuum concept can be viewed as rational concept rather than conceptual ones. This is the main purpose of this paper.

Any materials have three physical states as solid, liquid, and gas. Basically, the materials at some molecular cluster scales are the same for all the above physical states. So, a basic gauge tensor field can be attached to the basic molecular cluster elements (basic material element). By this kind of gauge field theory, different materials have different basic gauge fields.

Under this objective (material element) invariance principle, the motion concept of continuum in rational mechanics is expressed by the base vectors transformation [1]. Mathematically, the motion in continuum is a point-set transformation with the time as the implied parameter. (Unfortunately, in many researches (including some textbooks), it is treated as a coordinator transformation [2]. Although mathematically it is acceptable with some extent reservation, however, physically, the admissibility of physical motion and the material objective invariance principle are abandoned [3-5].) The geometrical theory based on the points-set transformation concept has been established by Chen Zhida, in 1987 [1]. He finds that the deformation tensor defined on coordinators embedded in the material is the looking-for motion transformation. Mathematically, Chen shows that the deformation tensor can be decomposed into the additive addition of a symmetrical tensor (represents intrinsic stretching of material element under discussion) and an orthogonal local rotation tensor [6-22] (represents intrinsic local rotation referring to its original orientation determined by neighboring materials). The researches exposes that the macro classical strain is expressed by addition of the intrinsic stretching of material element and its intrinsic local rotation.

Using Chen geometrical field theory of motion in continuum [3-22], the solid, liquid, and gas with fixed (invariant external) configuration have a common essential feature of zero classical strain (no macro deformation). Hence, the different phase states can be expressed by the different combination of intrinsic stretching and the local rotation. When the Chen decomposition form I and decomposition form II are used to meet the zero classical strain condition, there are three typical internal motion modes. Hence, it is very natural to find that: For zero classical strain (no macro deformation observed on its configuration surface, suitable container is required for liquid and gas to make up macro invariant configuration), the results show that: (1) For solid, the local rotation angular is zero. The material element has no intrinsic stretching. (2) For liquid, the local rotation will not change the basic gauge tensor. The material element has intrinsic plane stretching on the rotation plane. (3) For gas state, the intrinsic local rotation will amplify the basic gauge tensor. The material element has intrinsic stretching along the rotation direction.

Hence, for macro static continuum observed externally, under the understanding that no macro classical strain be observed, the material element has three different physical states: solid (no intrinsic stretching), fluid (plane intrinsic stretching), and gas (directional intrinsic stretching).

Then, a natural question will be asked: if the container is removed, what are the classical strains to be determined for different phase states. In this research, such a kind of condition is named as the three states. Rationally, the free conditions are defined by zero intrinsic stretching of material elements. Referring to this free condition, the stress concept is rationalized. Once these works have been done, the constitutive equations for the materials at multiple states are established. Based on the conventional standard, the classical constitutive equations must be included in this general unified formulation. This is certain for this research, as it is shown where it is appropriate.

The whole paper will be paragraphed as above. The reference papers [3-22] supply the additional documentation for further inquiring. For simplicity, the related results will be used directly.

## 2. Motion Transformation of Macro Static Continuum

For a material cluster element, embedding three coordinators $x^i, i=1,2,3$ with three basic vectors $\vec{g}_i^0$, then the material element is defined in continuum. Such a kind of coordinator system is named as commoving dragging coordinator system. Then the motion of the material element within unit time (suitably selected) is defined as:

$$\vec{g}_i = F_i^j \vec{g}_j^0 \tag{1}$$

Where, $\vec{g}_i$ is the current base vector, $F_j^i$ is the motion transformation. In mathematical sense, $F_j^i$ is a point-set transformation. In mechanics, it is named as deformation tensor. Locally, it is constructed by the displacement field $u^i$ (It is measured in initial coordinator system. Here, it is the displacement field of unit time; however, the velocity concept will not be used for the acceleration force is no discussed in the research. In fact, the zero acceleration means the static, here). The deformation tensor is defined as:

$$F_j^i = u^i\big|_j + \delta_j^i \tag{2}$$

Where, $u^i\big|_j$ represents the covariant-derivative, $\delta_j^i$ is a unit tensor.

Phenomenally, the static state of continuum is defined as no macro deformation is observed externally. That is the classical strain is zero. For different phase of the same material, its physical and mechanics implications are different.

**2.1 Static Solid**

Solid in natural environment has a fixed configuration. So, its static state is defined as the identical base transformation. Hence, the solid state is defined by the deformation tensor as:

$$F_j^i = \delta_j^i, \text{ for natural static solid} \tag{3}$$

As there is no displacement field, the classical strain is zero.

**2.2 Static Liquid**

For liquid, the material element has non-zero displacement field referring to its initial position. In natural static state, to make it has fixed configuration, a container is used to keep out its flow on horizontal plane. Using Chen decomposition form I, one has:

$$F_j^i = S_j^i + R_j^i \tag{4}$$

Where:

$$S_j^i = \frac{1}{2}(u^i\big|_j + u^j\big|_i) - (1 - \cos\Theta)L_k^i L_j^k \tag{5}$$

$$R_j^i = \delta_j^i + \sin\Theta \cdot L_j^i + (1 - \cos\Theta)L_k^i L_j^k \tag{6}$$

$$L_j^i = \frac{1}{2\sin\Theta}(u^i\big|_j - u^j\big|_i) \tag{7}$$

$$\sin\Theta = \frac{1}{2}[(u^1\big|_2 - u^2\big|_1)^2 + (u^2\big|_3 - u^3\big|_2)^2 + (u^3\big|_1 - u^1\big|_3)^2]^{\frac{1}{2}} \tag{8}$$

In above expressions, the parameter $\Theta$ ($-\pi/2 < \Theta < \pi/2$) is called local average rotation angel and tensor $L_j^k$ defines the local average rotation axis direction. $S_j^i$ is the intrinsic strain tensor, $R_j^i$ is an unit orthogonal rotation tensor. Not that the intrinsic strain and intrinsic local rotation are defined on material element referring to its initial configuration (the whole displacement and rigid rotation have no contribution to intrinsic deformation, which is eliminated by commoving feature of coordinator system).

As the classical strain is defined as:

$$\varepsilon_{ij} = \frac{1}{2}(u^i\big|_j + u^j\big|_i) \tag{9}$$

The static liquid will require that:
$$\varepsilon_{ij} \equiv 0 \tag{10}$$

So, the material intrinsic stretching will be:
$$S^i_j = -(1-\cos\Theta)L^i_k L^k_j = -(1-\cos\Theta)(L^i L_j - \delta^i_j) \tag{11}$$

Here, for simplicity, the following equations are used:
$$L_1 = L^1 = L^2_3 = -L^3_2, \quad L_2 = L^2 = L^3_1 = -L^1_3, \quad L_3 = L^3 = L^1_2 = -L^2_1 \tag{12}$$

Where, $L_i$ is the component of unit vector along the rotation axe direction.

For liquid in container, taken the $x^3$ on the free surface normal direction, as the flow direction is on ($x^1, x^2$) plane on statistical sense, the local rotation direction should be $L_3 = 1$, $L_1 = L_2 = 0$. Hence, the material element intrinsic stretching is obtained as:

$$S^1_1 = (1-\cos\Theta), \quad S^2_2 = (1-\cos\Theta), \quad S^3_3 = 0, \text{ others are zero} \tag{13}$$

It can be seen that the liquid has intrinsic isotropic expansion stretching on fluid surface plane direction. This will make the static liquid surface be an idea plane. Theoretically, as the surface normal direction has zero intrinsic stretching, the static liquid can have laminar flow structure, as it is observed in daily life. It represents our phenomenon experience well.

The local rotation will form steam line which will not be discussed in this paper.

**2.3 Static Gas**

For the gas, to keep it has fixed external configuration, a closed container must be used. For this case, the Chen decomposition form II should be used. By this formulation, he unit time deformation (deformation rate) is decomposed as:

$$F^i_j = \widetilde{S}^i_j + (\cos\theta)^{-1}\widetilde{R}^i_j \tag{14}$$

Where, the related items are:
$$\widetilde{S}^i_j = \frac{1}{2}(u^i\big|_j + u^j\big|_i) - (\frac{1}{\cos\theta} - 1)(\widetilde{L}^i_k \widetilde{L}^k_j + \delta^i_j) \tag{15}$$

$$(\cos\theta)^{-1}\widetilde{R}^i_j = \delta^i_j + \frac{\sin\theta}{\cos\theta}\widetilde{L}^i_j + (\frac{1}{\cos\theta} - 1)(\widetilde{L}^i_k \widetilde{L}^k_j + \delta^i_j) \tag{16}$$

$$\widetilde{L}^i_j = \frac{\cos\theta}{2\sin\theta}(u^i\big|_j - u^j\big|_i) \tag{17}$$

$$(\cos\theta)^{-2} = 1 + \frac{1}{4}[(u^1\big|_2 - u^2\big|_1)^2 + (u^2\big|_3 - u^3\big|_2)^2 + (u^3\big|_1 - u^1\big|_3)^2] \tag{18}$$

Here, the $\widetilde{S}^i_j$ is the intrinsic strain tensor, $\widetilde{R}^i_j$ is an unit orthogonal rotation tensor. Differing from the previous mode, the rotation direction tensor is changed into $\widetilde{L}^i_j$ with angular $\theta$ ($-\pi/2 < \theta < \pi/2$). Here, $u^i$ (previously, defined as displacement fields within the unit time duration) is the velocity field.

Comparing with the liquid, the important point is that: although the deformation tensor is continuous, the intrinsic strain and/or local rotation are not continuous. On mechanic sense, although the macro deformation is continuous, the intrinsic strain is discontinuous and the local rotation jumps from unit orthogonal rotation into orthogonal rotation with volume expansion (say bubbling in fluid environment or cracking in solid environment).

When zero classical strain condition is applied, the material intrinsic stretching will be:

$$\widetilde{S}_j^i = -(\frac{1}{\cos\theta} - 1)(\widetilde{L}_k^i \widetilde{L}_j^k + \delta_j^i) = -(\frac{1}{\cos\theta} - 1)\widetilde{L}^i \widetilde{L}_j \quad (19)$$

Where, $\widetilde{L}_1 = \widetilde{L}^1 = \widetilde{L}_3^2 = -\widetilde{L}_2^3$, $\widetilde{L}_2 = \widetilde{L}^2 = \widetilde{L}_1^3 = -\widetilde{L}_3^1$, $\widetilde{L}_3 = \widetilde{L}^3 = \widetilde{L}_2^1 = -\widetilde{L}_1^2$.

Near the container surface, the rotation direction should be along the surface normal direction. Taking this direction as the $\widetilde{L}_3 = 1$, $\widetilde{L}_1 = \widetilde{L}_2 = 0$, the intrinsic stretching is:

$$\widetilde{S}_3^3 = -(\frac{1}{\cos\theta} - 1), \text{ others are zero} \quad (20)$$

It shows that the material element is compressed by the container boundary.

On the other hand, the expansion local rotation gives out an isotropic expansion $(\frac{1}{\cos\theta} - 1)\delta_{ij}$, so the net expansion is parallel to the container surface. The internal materials isotropic expansion is the intrinsic features of gas. In classical statistic theory, the boundary bounce mode is used. This logically has no contract with the boundary compressing here.

Summing above results: (1) the solid has no intrinsic deformation; (2) the liquid has intrinsic plane expansion on local rotation plane, the local rotation is an unit orthogonal rotation; (3) the gas has intrinsic compressive on its local rotation direction, the local rotation is combined with isotropic expansion. Generally, the intrinsic stretching will change the shape of material elements.

## 3. Motion Transformation of Intrinsic Free Continuum

For a free material element in continuum, it should has no intrinsic stretching. Its orientation is determined by the continuum as whole. If in statistical sense the material elements are in free state, the continuum is defined as intrinsic free continuum. Based on previous formulation, the intrinsic free continuum is defined as:

$$S_j^i = 0, \text{ or } \widetilde{S}_j^i = 0 \quad (21)$$

This means that the classical strain may be non-zero, hence macro displacement field or deformation can be observed.

### 3.1 Free Solid

For free solid, the deformation is Equation (3). It is an identical motion transformation. That means the free solid material elements are fixed in continuum. In fact, this picture is overwhelmingly adopted in many textbooks as the definition of continuum in mechanics. By this definition, the classical strain for free solid is:

$$F_j^i = \delta_j^i, \quad \varepsilon_{ij} = 0, \text{ free solid} \quad (22)$$

Comparing with previous results, the static solid is free solid. This is the special feature of solid. It means that the solid material has no motion without external force action.

### 3.2 Free Liquid

For liquid, its free state is defined by the deformation tensor:

$$F_j^i = R_j^i = \delta_j^i + \sin\Theta \cdot L_j^i + (1 - \cos\Theta)L_k^i L_j^k \quad (23)$$

It is a pure local rotation. The gauge tensor of the material element is invariant. That is:

$$g_{ij} = \vec{g}_i \cdot \vec{g}_j = R_i^k R_j^l \vec{g}_k^0 \vec{g}_l^0 = R_i^k R_j^l g_{kl}^0 \quad (24)$$

For spatial isotropic material element, $g_{ij}^0 = g^0 \delta_{ij}$, one has: $g_{ij} = g^0 \delta_{ij} = g_{ij}^0$. Hence, the liquid material element is gauge field invariant. The free state is defined by the local rotation (orientation variation). The local rotation is limited by the continuum feature as a whole, as the macro classical strain is not zero. By Equation (5),, one has:

$$\varepsilon_{ij} = \frac{1}{2}(u^i\big|_j + u^j\big|_i) = (1-\cos\Theta)L_k^i L_j^k \quad (25)$$

For the local rotation along $L_3 = 1$ direction (see Equation (12)), the classical strain components are:

$$\varepsilon_{11} = -(1-\cos\Theta), \quad \varepsilon_{22} = -(1-\cos\Theta), \text{ others are zero} \quad (26)$$

Phenomenally, the continuum supplies the essential compressing to keep the liquid element has invariant gauge. So, the free liquid will form flow-stream line on the rotational plane.

If the macro mechanical feature of liquid is expressed by viscosity parameters ($\tilde{\lambda}, \tilde{\mu}$), the internal initial stress defined by the free state of continuum is:

$$\sigma_{11} = \sigma_{22} = -2(\tilde{\lambda}+\tilde{\mu})(1-\cos\Theta), \quad \sigma_{33} = -2\tilde{\lambda}(1-\cos\Theta), \text{ others are zero} \quad (27)$$

It is an anisotropy stress field. Generally speaking, it will drive the liquid element drifts along the rotation direction while rotating along the direction. This phenomenon is typical in daily life as the high viscous fluid and low viscous fluid are separated by itself in free continuum.

Note that, for liquid, the isotropic stress components can be defined as:

$$-p_0 \delta_{ij} = -2\tilde{\lambda}(1-\cos\Theta)\delta_{ij} \quad (28)$$

So, the rational definition of static pressure of fluid is:

$$p_0 = 2\tilde{\lambda}(1-\cos\Theta) \quad (29)$$

It shows that the static pressure of fluid is determined by the fluid viscosity $\tilde{\lambda}$ and its local rotation angle. Generally speaking, for the same material composition, the bigger is the local rotation angle, the higher pressure is.

The surface tension of liquid is rationally defined as:

$$\sigma_{ij} = -2\tilde{\mu}(1-\cos\Theta)\delta_{ij}, \quad i,j=1,2 \text{ on surface} \quad (30)$$

Note that plane stress on rotational plane has no contribution to the liquid pressure as the rotation has random distribution. It shows that the free liquid material elements have an intrinsic surface contraction force to keep its shape. For given material features, by measuring the surface tension, the $\Theta$ parameter can be measured (generally, for most fluid, the $\tilde{\mu}$ can be well measured). In fact, the Equations (29) and (30) can be used to determine the parameter $\tilde{\lambda}$. This parameter is omitted in most fluid mechanics textbook. This problem is highly criticized (Lodge, A.S, 1974, [23]).

In fact, by the classical strain Equation (25), the general form of stress can also be written as:

$$\sigma_{ij} = -2(\tilde{\lambda}+\tilde{\mu})(1-\cos\Theta)\delta_{ij} + 2\tilde{\mu}(1-\cos\Theta)L_i L_j \quad (31)$$

For a small liquid drop, the isotropic compressive stress is interpreted as the atmosphere pressure, the second item on the right can be interpreted as the liquid drop expansion on the surface direction. Hence, the liquid drop can have a fixed shape with an appropriate highness. This equation can found its usage in many cases.

### 3.3 Free Gas

Free gas is defined by the motion transformation:

$$F_j^i = (\cos\theta)^{-1}\widetilde{R}_j^i = \delta_j^i + \frac{\sin\theta}{\cos\theta}\widetilde{L}_j^i + (\frac{1}{\cos\theta}-1)(\widetilde{L}_k^i\widetilde{L}_j^k + \delta_j^i) \qquad (32)$$

If the initial gauge tensor is selected as: $g_{ij}^0 = g^0\delta_{ij}$, the current gauge tensor is: $g_{ij} = (\frac{1}{\cos\theta})^2 g^0\delta_{ij}$. So, within unit time, the length gauge is increased by a factor $1/\cos\theta$. In conventional nomination, this factor is named as expansion rate of gas. It depends on the internal motion of gas as a continuum. This is represented by the local rotation angle $\theta$. Unlike the free liquid (invariant gauge tensor), free gas material will expansion infinitely. So, a closed container is needed to keep them in a configuration.

For free gas in continuum mechanics, the classical strain is:

$$\varepsilon_{ij} = \left(\frac{1}{\cos\theta} - 1\right)\widetilde{L}_i\widetilde{L}_j \qquad (33)$$

When the open surface normal direction is taken as the ($\widetilde{L}_3 = 1$) local rotation direction, the classical strain is:

$$\varepsilon_{33} = \left(\frac{1}{\cos\theta} - 1\right) \qquad (34)$$

It means that the gas will escape from the surface. As a simple example, if a rectangular container is used, when the $x^3 = H$ plane ($x^3 = 0$ is the opposite boundary surface) is opened suddenly, the escape velocity distribution for the gas material within the container will be approximated as (in statistical sense): $u_3 = \left(\frac{1}{\cos\theta} - 1\right) \cdot x^3$.

The classical stress for free gas material element is:

$$\sigma_{ij} = \widetilde{\lambda}(\frac{1}{\cos\theta}-1)\delta_{ij} + 2\widetilde{\mu}(\frac{1}{\cos\theta}-1)\widetilde{L}_i\widetilde{L}_j \qquad (35)$$

The first item on the right is an isotropy expansion pressure, it has the opposite sign referring to the liquid. Hence, in rational mechanics, the gas pressure should be defined as:

$$P_0 = \widetilde{\lambda}(\frac{1}{\cos\theta}-1) \qquad (36)$$

Note that the second item has no contribution to the gas pressure as the rotation has random distribution.

Summering up above results: (1) continuum composed by intrinsic free solid has no classical strain; (2) continuum (fluid) composed by intrinsic free liquid has a static compressive pressure; (3) continuum (gas or vapor) composed by intrinsic free gas has a static expansion pressure.

Note that, although at the free state the material has classical strain, its macro statistical value is zero for random orientation distribution. Therefore, for idea continuum, the static state is equivalent with the free state. However, for modern industry, in many cases, the liquid or gas may have coherent orientation distribution. In these cases, the static state and the free state should be identified independently. Strictly speaking, the static state is defined externally as no configuration variation. The free state is defined internally (physically) as no molecular cluster scale deformation (intrinsic stretching).

**4. Constitutive Equations for Continuum**

For continuum, viewing it from physics, the free states (zero intrinsic strain) should be taken as the reference state. Viewing from mechanics deformation theory, the static state should be taken as the reference configuration. Except for solid, the incremental stress reference and the incremental strain reference are different.

In Green stress definition (which is usually used in physics), the intrinsic stretching stress is defined as:

$$\sigma^i_j = \lambda(F^l_l - 3)\delta_{ij} + 2\mu(F^i_j - \delta^i_j) \tag{37-1}$$

In Chen rational mechanics, the classical stress is defined as:

$$\sigma^i_j = \lambda S^l_l \delta^i_j + 2\mu S^i_j \tag{37-2}$$

Where, the parameters ($\lambda, \mu$) is determined by the physical features of materials. This equation will be used to derive the classical constitutive equations.

Chen Zhida argued that the local rotation has no contribution to the stress. He pointed out that the classical strain in the constitutive definition should be replaced by the intrinsic strain (intrinsic stretching). In this paper, the related problems will be fully discussed from physical and statistical points of view.

For continuum, as the local rotation direction (attached on the material element) is random for general cases, they contribute zero stress to macro continuum stress observed externally. So, on statistical sense, the macro stress of continuum is determined by the intrinsic stretching tensor (intrinsic strain tensor). That is:

$$\sigma^i_j = \frac{1}{A_\Sigma} \oint_\Sigma \sigma^i_j d\Sigma = \lambda S^l_l \delta^i_j + 2\mu S^i_j \tag{38}$$

The surface integration expression is used to emphasize the statistical sense. It shows that, for stress, the reference state of material should be its free state. For liquid and gas at free states, the initial strain is has zero average. However, the initial stress is not zero.

For infinitesimal incremental deformation, the classical strain is defined as:

$$e_{ij} = \frac{1}{2}(\frac{\partial U^i}{\partial X^j} + \frac{\partial U^j}{\partial X^i}) \tag{38}$$

Here, the laboratory coordinator system ($X^1, X^2, X^3$) is used. $U^i$ is the spatial macro displacement within the unit time under discussion. The classical strain attached on the material element has zero contribution to the average, because their rotation direction and orientation have randomly distribution.

In deformation mechanics point of views, the natural continuum which can be taken as a reference should be: zero statistical classical strain reference.

### 4.1 Solid Continuum

For solid continuum, referring to its free state, the classical constitutive equation is:

$$\sigma_{ij} = \lambda(e_{ll})\delta_{ij} + 2\mu e_{ij} \tag{39}$$

### 4.2 Liquid Continuum

For liquid continuum, its free state is defined as:

$$F^i_j = R^i_j(\Theta) \tag{40}$$

For infinitesimal deformation, the incremental deformation is assumed to have no contribution to the intrinsic local rotation of material element. Hence, the stress is:

$$\sigma_{ij} = -2(\tilde{\lambda} + \tilde{\mu})(1 - \cos\Theta)\delta_{ij} + 2\tilde{\mu}(1 - \cos\Theta)L_i L_j + \tilde{\lambda}(e_{ll})\delta_{ij} + 2\tilde{\mu}e_{ij} \tag{41}$$

For random orientation distribution, it is simplified as:

$$\sigma_{ij} = -2(\tilde{\lambda} + \tilde{\mu})(1-\cos\Theta)\delta_{ij} + \tilde{\lambda}(e_{ll})\delta_{ij} + 2\tilde{\mu}e_{ij} \tag{42}$$

Or, in conventional form:

$$\sigma_{ij} = -p_0\delta_{ij} + \tilde{\lambda}(e_{ll})\delta_{ij} + 2\tilde{\mu}e_{ij} \tag{43}$$

It is clear, if the contribution from incremental local rotation is included, the static pressure should be modified. As the static pressure is temperature dependent, it can be inferred that the local rotation angle $\Theta$ plays the similar role as the temperature parameter.

For many fluid, the item $\tilde{\lambda}(e_{ll})\delta_{ij}$ is dropped under the names "incompressible materials" ($e_{ll} = 0$) or "inelastic materials" ($\tilde{\lambda} \ll \tilde{\mu}$).

**4.3 Gas Continuum**

For gas continuum, its free state is defined as:

$$F_j^i = \frac{1}{\cos\theta}\tilde{R}_j^i(\theta) \tag{44}$$

For infinitesimal deformation, the incremental deformation is assumed to have no contribution to the intrinsic local rotation of material element. Hence, the stress is:

$$\sigma_{ij} = \tilde{\lambda}(\frac{1}{\cos\theta}-1)\delta_{ij} + 2\tilde{\mu}(\frac{1}{\cos\theta}-1)\tilde{L}_i\tilde{L}_j + \tilde{\lambda}(e_{ll})\delta_{ij} + 2\tilde{\mu}e_{ij} \tag{45}$$

Or, in conventional form for random orientation distribution:

$$\sigma_{ij} = P_0\delta_{ij} + \tilde{\lambda}(e_{ll})\delta_{ij} + 2\tilde{\mu}e_{ij} \tag{46}$$

It is clear, if the contribution from incremental local rotation is included, the static pressure should be modified. As the static pressure is temperature dependent, it can be inferred that the local rotation angle $\theta$ plays the similar role as the temperature parameter.

For many gas continuums, the shear is omitted, and the constitutive equation becomes:

$$\sigma_{ij} = P_0\delta_{ij} + \tilde{k}(e_{ll})\delta_{ij} \tag{47}$$

Here, the bulk compressibility parameter $\tilde{k}$ is used to replace the viscosity parameter $\tilde{\lambda}$ as their physical implications are different (although they are almost equal in value).

In thermal mechanics, the equation is written as:

$$P = P_0 + \tilde{k}dV \tag{48}$$

It is clear that, the overwhelming simplification of constitutive equation of gas continuum cuts its link with the deformation mechanics sharply. No doubt, this empirical altitude will damage the fully development of gas dynamics.

Summering above results, the unified constitutive equations for continuum are

$$\sigma_j^i = \lambda S_l^l \delta_j^i + 2\mu S_j^i, \text{ Chen decomposition form I} \tag{49-1}$$

$$\sigma_j^i = \lambda \tilde{S}_l^l \delta_j^i + 2\tilde{\mu} S_j^i, \text{ Chen decomposition form II} \tag{49-2}$$

By these equations, the stress is the classical stress. When the continuity of stress concept is used in deformation theory, the stress defined above is continuous, although the deformation may be not continuous.

Here, it should be pointed out that the local rotation to another kind of stress which is related with the internal curvature of continuum. It is related with the micro-deformation of continuum. As a first approximation, it is related with thermo mechanics. This topic will be expanded bellow.

**5. Phase Transition**

The above formulation is about the deformation within an unit time. If the unit time is much longer than the characteristic time of liquid (composing fluid) or gas (composing vapor), the formulation must be modified. For engineering practice, the continuum is assumed to be stable within the unit time. In fact, it is the usually case. However, this does not mean the intrinsic motion of the basic material element has long characteristic time.

In fact, introducing a characteristic time scale $\tau$ will be useful.

Firstly, the within the fraction time $\tau \approx \frac{1}{N} \ll 1$ ($\tilde{\tau} < \tau$ for gas) duration, the local rotation angle will be expressed:

$$\Theta = \frac{\Theta_\tau}{\tau} \tag{50-1}$$

$$\theta = \frac{\theta_\tau}{\tilde{\tau}} \tag{50-2}$$

They are interpreted as the local rotation angle within time $\tau$. To meet our formulation, the following conditions are applied:

$$0 \leq \Theta = \frac{\Theta_\tau}{\tau} < \pi/2, \quad 0 \leq \theta = \frac{\theta_\tau}{\tilde{\tau}} < \pi/2 \tag{51}$$

Where, the negative sign is dropped as the absolute value will be discussed bellow.

Hence, to modify the local rotation angle definition, on statistical sense, the best way is to express the free liquid deformation as:

$$F_j^i = R_j^i(\frac{\Theta_\tau}{\tau}) \tag{52}$$

The similar procedure is applied to the free gas.

$$F_j^i = \frac{1}{\cos(\frac{\theta_\tau}{\tilde{\tau}})} \tilde{R}_j^i(\frac{\theta_\tau}{\tilde{\tau}}) \tag{53}$$

By this way, the $\Theta$ and $\theta$ still is defined by unit time duration.

For simplicity, in the following discussion, the $\tau = 1$ is taking as the unite time duration. The above discussion is to emphasize the statistical sense for the related formulations.

Viewing the local rotation as a stochastic process:

$$\Theta = \Theta_0 + \delta\Theta \tag{54-1}$$

$$\theta = \theta_0 + \delta\theta \tag{54-2}$$

For simplicity, it is viewed as a Brownian motion with normal distribution.

Under this understanding, the local rotation angle is represented as a stochastic process defined by the characteristic function of normal distribution:

$$f_\Theta(t) = \exp\left(\tilde{i}\,\Theta_0 \cdot t - \frac{1}{2}\sigma_\Theta^2 \cdot t^2\right) \quad (55\text{-}1)$$

$$f_\theta(t) = \exp\left(\tilde{i}\,\theta_0 \cdot t - \frac{1}{2}\sigma_\theta^2 \cdot t^2\right) \quad (55\text{-}2)$$

Where, $\tilde{i} = \sqrt{-1}$ is the sign of imaginary number, $\Theta_0$ and $\theta_0$ are the mean value of $\Theta$ and $\theta$ process respectively, $\sigma_\Theta^2$ and $\sigma_\theta^2$ are their variance. The corresponding probability distribution function is:

$$\tilde{P}(\Theta) = \frac{1}{\sqrt{2\pi} \cdot \sigma_\Theta} \cdot \exp\left[-\frac{(\Theta - \Theta_0)^2}{2\sigma_\Theta^2}\right] \quad (56\text{-}1)$$

$$\tilde{P}(\theta) = \frac{1}{\sqrt{2\pi} \cdot \sigma_\theta} \cdot \exp\left[-\frac{(\theta - \theta_0)^2}{2\sigma_\theta^2}\right] \quad (56\text{-}2)$$

Based on above formulations, using the statistical mechanics theory, the temperature incremental $dT$ is related with the variance as:

$$\sigma_\Theta^2 = C_\Theta \cdot \frac{dT}{\rho} \quad (57\text{-}1)$$

$$\sigma_\theta^2 = C_\theta \cdot \frac{dT}{\rho} \quad (57\text{-}2)$$

Here, the continuum mass density $\rho$ is invariant for phase transition of isochoric process. $C_\Theta$ and $C_\theta$ are material thermal parameters. For phase transition discussion, it can be viewed as a constant (in fact, they should be determined by statistical physics for different micro dynamics processes).

With above preliminary preparation, the phase transition problem can be discussed.

**5.1 Solid-Gas Transition**

For solid, its intrinsic characteristic time is very large. For unit time duration, its average intrinsic local rotation is almost zero. For solid continuum, when the local rotation variance $\sigma_\Theta$ is bigger than a critical angle $\Theta_S$ (which is determined by material features), a fraction of solid will be cracked out from its surroundings. The cracked out materials will be free at low pressure environment. So, they are viewed as gas here.

For isotropic cracking process with certain possibility (defined by variance level $\sigma_\Theta$), initially, the solid deformation is:

$$F_j^i = S \cdot \delta_j^i + R_j^i(\sigma_\Theta) \quad (58)$$

It produces the stress which is:

$$\sigma_{ij} = -2(\lambda + \mu)(1 - \cos\sigma_\Theta)\delta_{ij} + 2\mu(1 - \cos\sigma_\Theta)L_i L_j + (3\lambda + 2\mu)S \cdot \delta_{ij} \quad (59)$$

For homogenous distribution of rotation direction ($mean(L_i L_j) = \frac{1}{3}$, here after), the isotropic stress is:

$$\sigma = -2(\lambda + \mu)(1 - \cos\sigma_\Theta) + \frac{2\mu}{3}(1 - \cos\sigma_\Theta) + (3\lambda + 2\mu)S \quad (60)$$

After cracking, for the cracked out materials, the gas deformation is:

$$F_j^i = \frac{1}{\cos\theta_0}\tilde{R}_j^i(\theta_0) \quad (61)$$

Similarly, for homogenous distribution of rotation direction ($mean(\tilde{L}_i \tilde{L}_j) = \frac{1}{3}$, hear, after), its corresponding isotropic stress is:

$$\sigma_{ij} = (\tilde{\lambda} + \frac{2}{3}\tilde{\mu}) \cdot (\frac{1}{\cos\theta_0} - 1)\delta_{ij} \tag{62}$$

The macro volume continuity gives out the condition equation:

$$\frac{1}{\cos\theta_0} - 1 = S \tag{63}$$

For solid-gas transition, the isotropic stress should be continuous. So, the transition equation is:

$$(3\lambda + 2\mu)(\frac{1}{\cos\theta_0} - 1) - 2(\lambda + \frac{2}{3}\mu)(1 - \cos\sigma_\Theta) = (\tilde{\lambda} + \frac{2}{3}\tilde{\mu}) \cdot (\frac{1}{\cos\theta_0} - 1) = P_G \tag{64}$$

For the solid-gas transition, the $P_G$ pressure is the measured externally. For low pressure, the $\theta_0$ is small. When the $P_G$ is taken as the variable, the $\theta_0$ is determined by the last equality equation.

$$(\frac{1}{\cos\theta_0} - 1) = \frac{P_G}{\tilde{\lambda} + \frac{2}{3}\tilde{\mu}} \tag{65}$$

As the gas local rotation $\theta_0$ is temperature dependent, so the global temperature is determined by this equation, also.

For given material parameters, putting Equation (57-1) into Equation (64), the incremental temperature is determined by the equation:

$$1 - \cos\sqrt{C_\Theta \frac{dT}{\rho_S}} = \frac{P_G}{2(\lambda + \frac{2}{3}\mu)} \cdot (\frac{3\lambda + 2\mu}{\tilde{\lambda} + \frac{2}{3}\tilde{\mu}} - 1) \tag{66}$$

Here, the mass density is defined for solid state. Hence, the $dT$ is determined.

The maximum allowable solid-gas transition pressure range is given by the equation:

$$0 \leq \frac{P_G}{2(\lambda + \frac{2}{3}\mu)} \cdot (\frac{3\lambda + 2\mu}{\tilde{\lambda} + \frac{2}{3}\tilde{\mu}} - 1) \leq 1 \tag{67}$$

By instinct, the critical temperature for the triple point of solid-liquid-gas is defined by the pressure:

$$P_{triple} = \frac{2(\lambda + \frac{2}{3}\mu)}{\frac{3\lambda + 2\mu}{\tilde{\lambda} + \frac{2}{3}\tilde{\mu}} - 1} \tag{68}$$

It is completely determined by the solid features and gas features. The triple temperature can be determined, indirectly, by the equation:

$$(\frac{1}{\cos\theta(T_{triple})} - 1) = \frac{P_{triple}}{\tilde{\lambda} + \frac{2}{3}\tilde{\mu}} = \frac{2(\lambda + \frac{2}{3}\mu)}{(3\lambda + 2\mu) - (\tilde{\lambda} + \frac{2}{3}\tilde{\mu})} \tag{69}$$

It is completely determined by the solid features and gas features, also.

### 5.2 Liquid-Gas Transition

For liquid-gas process with certain possibility (defined by variance level $\sigma_\Theta$), the liquid deformation tensor is:

$$F_j^i = R_j^i(\Theta_0 + \sigma_\Theta) \tag{70}$$

For such a liquid, its initial pressure is defined as:

$$p_0 = 2(\tilde{\lambda}_L + \frac{2}{3}\tilde{\mu}_L)(1-\cos\Theta_0) \tag{71}$$

Here and after, the liquid parameters are labeled with L to distinguish from gas.

Based on phase transition curves, the following condition is met:

$$2(\tilde{\lambda}_L + \frac{2}{3}\tilde{\mu}_L)(1-\cos\Theta_0) \geq P_{triple} \tag{72}$$

At the liquid-gas transition process, the liquid stress tensor is:

$$\sigma_{ij} = -2(\tilde{\lambda}_L + \frac{2}{3}\tilde{\mu}_L)[1-\cos(\Theta_0 + \sigma_\Theta)]\delta_{ij} \tag{73}$$

Hence, using gas Equations (61) and (62), the stress continuity condition is:

$$2(\tilde{\lambda}_L + \frac{2}{3}\tilde{\mu}_L)[1-\cos(\Theta_0 + \sigma_\Theta)] = (\tilde{\lambda} + \frac{2}{3}\tilde{\mu})(\frac{1}{\cos\theta_0} - 1) = P_G \tag{74}$$

It means that the gas expansion stress is balance by the liquid compressive stress. (In solid-gas transition, the gas expansion stress is supplied by the solid volume expansion force. This can be true only for low temperature case.) It is clear that, here, $P_G > P_{triple}$.

Using the $P_G$ as the variables, the transition temperature incremental is determined by equation:

$$1-\cos(\Theta_0 + \sqrt{C_\Theta \frac{dT}{\rho_L}}) = \frac{P_G}{2(\tilde{\lambda}_L + \frac{2}{3}\tilde{\mu}_L)} \tag{75}$$

The global temperature is determined by the equation (indirectly):

$$\frac{1}{\cos\theta_0(T)} - 1 = \frac{P_G}{\tilde{\lambda} + \frac{2}{3}\tilde{\mu}} \tag{76}$$

Generally speaking, as $P_G > P_{triple}$, the temperature is higher than the solid-gas transition temperature.

**5.3 Solid-Liquid Transition**

For many solid at medium temperature and pressure, the stochastic local rotation is not zero-mean. For example, when some kinds of coherent patterns happened in the solid materials, the shear stress will crack the materials into several pieces. Hence, the mean local rotation is a finite value $\Theta_S$. This feature is expressed by the deformation tensor:

$$F^i_j = R^i_j(\Theta_S + \sigma_\Theta) \tag{77}$$

Here, the mean value of local rotation angle is determined by the yield stress of solid as:

$$2(\lambda + \frac{2}{3}\mu)(1-\cos\Theta_S) = \sigma_{yield} \tag{78}$$

For liquid, the deformation tensor is:

$$F^i_j = R^i_j(\Theta_L) \tag{79}$$

The stress continuity equation is:

$$2(\lambda + \frac{2}{3}\mu)(1-\cos(\Theta_S + \sigma_\Theta)] = 2(\tilde{\lambda}_L + \frac{2}{3}\tilde{\mu}_L)(1-\cos\Theta_L) = P_L \tag{80}$$

As solid parameters ($\lambda, \mu$) are much larger than liquid parameters ($\tilde{\lambda}_L, \tilde{\mu}_L$), the $P_L$ will be at the range:

$$P_{triple} \leq \sigma_{yield} \leq P_L \tag{81}$$

The global temperature is indirectly determined by equation:

$$2(\tilde{\lambda}_L + \frac{2}{3}\tilde{\mu}_L)[1 - \cos\Theta_L(T)] = P_L \tag{82}$$

Generally speaking, although $P_L > P_G$, as liquid parameters ($\tilde{\lambda}_L, \tilde{\mu}_L$) are much larger than gas parameters ($\tilde{\lambda}, \tilde{\mu}$), the temperature is lower than the liquid-gas transition temperature. This can be inferred from the related equations.

Using the $P_L$ as the variables, the transition temperature incremental is determined by equation:

$$2(\lambda + \frac{2}{3}\mu)[1 - \cos(\Theta_S + \sqrt{C_\Theta \frac{dT}{\rho_S}})] = P_L \tag{83}$$

Now, it is time to turn to the supercritical fluid region.

**5.4 Super-Fluid Transition**

For liquid materials, its deformation is:

$$F_j^i = R_j^i(\Theta) \tag{84}$$

For gas, the deformation is:

$$F_j^i = \frac{1}{\cos\theta}\tilde{R}_j^i(\theta) \tag{85}$$

By the geometrical Equations (6) and (16), when $\Theta \to \pi/2$, or $\theta \to \pi/2$, the materials cannot be viewed as continuum, because there are no continuous displacement field (here, it is velocity) for such cases. Observing equation:

$$2(\tilde{\lambda}_L + \frac{2}{3}\tilde{\mu}_L)(1 - \cos\Theta) = (\tilde{\lambda} + \frac{2}{3}\tilde{\mu})(\frac{1}{\cos\theta} - 1) = P_C \tag{86}$$

For $\Theta \to \pi/2$, the liquid does no form a continuum in conventional sense. The critical pressure is defined as:

$$2(\tilde{\lambda}_L + \frac{2}{3}\tilde{\mu}_L) = (\tilde{\lambda} + \frac{2}{3}\tilde{\mu})(\frac{1}{\cos\theta(T_C)} - 1) = P_C \tag{87}$$

For $\theta \to \pi/2$, the pressure tends to infinite. That is:

$$P = (\tilde{\lambda} + \frac{2}{3}\tilde{\mu})(\frac{1}{\cos\theta(T)} - 1) \to \infty, \quad \text{when} \quad \theta \to \pi/2 \tag{87}$$

Hence, the super-fluid region is defined by condition equations:

$$P \geq P_C = 2(\tilde{\lambda}_L + \frac{2}{3}\tilde{\mu}_L), \quad T \geq T_C \tag{88}$$

It shows that, the critical pressure is completely determined by the fluid parameters. Where, the critical temperature is determined (indirectly) by equation:

$$\frac{1}{\cos\theta(T_C)} = 1 + \frac{2(\tilde{\lambda}_L + \frac{2}{3}\tilde{\mu}_L)}{(\tilde{\lambda} + \frac{2}{3}\tilde{\mu})} \tag{89}$$

It shows that, the critical temperature is determined by the ratio of liquid parameter over gas parameter.

Summering above results: the phase transition phenomenon is well formulated by the unique physical requirements stress continuity (pressure continuity) for continuous temperature variation. The typical phase diagram of continuum be obtained by the related equations given in this paper.

## 6. Application: Rock as Multiphase Continuum

There are two different multiphase continuum definitions. The multiphase continuum in physics is a topic for statistical mechanics, where the materials have the same composition. In this case, the internal interaction is temperature dependent, and the dynamics in a scale less than molecular cluster scale is concerned. It is too complicated for deformation mechanics. This topic will not be discussed here further.

In mechanical engineering, the multiphase continuum is referring to the continuum composed by solid, liquid, and gas. They usually have different composition. The solid forms the volume frame, while the liquid and gas are filled in the volume. They have contribution to stress, but usually has no direct contribution to volume variation.

For simplicity, taking the volume variant deformation as an example, the macro deformation tensor is:

$$F_j^i = \gamma(1+e)\delta_j^i + \alpha R_j^i(\Theta) + \beta \frac{1}{\cos\theta} \widetilde{R}_j^i(\theta) \tag{90}$$

Where, the solid, liquid, and gas contribution coefficients are $\gamma, \alpha,$ and $\beta$. Their total should be unit one. $\gamma + \alpha + \beta = 1$. Then, the stress field is:

$$\sigma_{ij} = [\gamma(3\lambda + 2\mu)\cdot e - 2\alpha(\widetilde{\lambda}_L + \frac{2}{3}\widetilde{\mu}_L)(1 - \cos\Theta) + \beta(\widetilde{\lambda} + \frac{2}{3}\widetilde{\mu})(\frac{1}{\cos\theta} - 1)]\delta_{ij} \tag{91}$$

The static state, which is taken as the reference configuration is defined by the deformation:

$$F_j^i = \gamma\delta_j^i + \alpha R_j^i(\Theta_0) + \beta \frac{1}{\cos\theta_0} \widetilde{R}_j^i(\theta_0) \tag{92}$$

The static state stress is:

$$\sigma_{ij}^0 = [-2\alpha(\widetilde{\lambda}_L + \frac{2}{3}\widetilde{\mu}_L)(1 - \cos\Theta_0) + \beta(\widetilde{\lambda} + \frac{2}{3}\widetilde{\mu})(\frac{1}{\cos\theta_0} - 1)]\delta_{ij} \tag{93}$$

Based on the geometrical meaning of each items, in Chen rational mechanics, the length variation is:

$$\delta g = e \tag{94}$$

The stress variation is:

$$\delta\sigma_{ij} = [\gamma(3\lambda + 2\mu)\cdot e - 2\alpha(\widetilde{\lambda}_L + \frac{2}{3}\widetilde{\mu}_L)(\cos\Theta - \cos\Theta_0) + \beta(\widetilde{\lambda} + \frac{2}{3}\widetilde{\mu})(\frac{1}{\cos\theta} - \frac{1}{\cos\theta_0})]\delta_{ij} \tag{95}$$

This equation can be simplified as:

$$\delta\sigma_{ij} = [\gamma(3\lambda + 2\mu)\cdot e + 2\alpha(\widetilde{\lambda}_L + \frac{2}{3}\widetilde{\mu}_L)(\sin\Theta_0)\delta\Theta + \beta(\widetilde{\lambda} + \frac{2}{3}\widetilde{\mu})(\frac{\sin\theta_0}{\cos^2\theta_0})\delta\theta]\delta_{ij} \tag{97}$$

Hence, if one define the stress by the effective parameters ($\bar{\lambda}, \bar{\mu}$), then the stress is expressed as:

$$\delta\sigma_{ij} = (3\bar{\lambda} + 2\bar{\mu})\cdot e \tag{98}$$

By comparing the Equations (97) and (98), one has:

$$3\bar{\lambda} + 2\bar{\mu} = \gamma\cdot(3\lambda + 2\mu) + 2\alpha(\widetilde{\lambda}_L + \frac{2}{3}\widetilde{\mu}_L)(\sin\Theta_0)\frac{\delta\Theta}{e} + \beta(\widetilde{\lambda} + \frac{2}{3}\widetilde{\mu})(\frac{\sin\theta_0}{\cos^2\theta_0})\cdot\frac{\delta\theta}{e} \tag{100}$$

It says the effective mechanic parameters are the weight sum of solid, liquid, and gas features. Generally speaking, the $\delta\Theta/e$ and $\delta\theta/e$ are material feature constants. In mining industry, the infinitive effective mechanical parameter means rock bursting.

For gas driven bursting in mining industry, the condition can be predicted by $\theta_0 \geq \theta_{gas}$, where the $\theta_{gas}$ can be measured by laboratory experiments. For $\Theta_0 \to \pi/2$, water driven bursting is produced. Hence, by measuring the compressive parameters of rock, the potential rock bursting can be predicted.

## 7. Conclusions

In this research based on Chen Rational Mechanics frame, the motion of mater in continuum sense is described by the base vector transformation: $\vec{g}_j(t,x) = F_j^i(t,x) \cdot \vec{g}_i^0(x)$ (the spatial and time parameters will be dropped, here after). Once an arbitral initial coordinator is selected, a material element will be labeled by this coordinator combining with local initial base vector (on this sense, the coordinator and base vectors are embedded into the continuum), no matter how its current configuration is deformed. By this way, omitting the global translation and rotation of the continuum as a whole, the matter motion in continuum is purely the base vector intrinsic stretching and local relative intrinsic rotation referring to its initial configuration (the should be configuration determined by the continuum motion surrounding the material element under discussion). In observation sense, the commoving dragging coordinator system (defined by coordinators and base vectors) variation is the natural results of matter motion as a continuum. Hence, such a geometrical field theory is a natural selection for continuum motion. (If coordinator transformation method is used, there are too many jobs putting on the relation between the $x^i$ coordinator system and the laboratory $X^i$ coordinator system. Then, the true physical essential variation of gauge tensor is waiting to be recovered, which is not practical for complicated deformation. Furthermore, the material objective invariance may not be met. Although for infinitesimal deformation this is not serious, it do cause serious error for large deformation or deformation with local rotation.)

In such a kind of natural geometrical field selection, the static solid is defined by: $F_j^i = \delta_j^i$. It is an identical transformation.

For static liquid, within unit time, the basic material element motion is defined as: $F_j^i = R_j^i(\Theta)$. It means that, the liquid material has a local relative rotation referring to its "should be" configuration unit-time-before. For continuum composed by liquid, it has a classical expansion on its rotation plane. Hence, to make the continuum has a fixed external configuration, a container should be used (top open boundary under gravity field). As the local rotation has no classical strain on rotation direction, a natural laminar structure (each layer has different local rotation angular) may be formed. This is the basic feature for static continuum. By this sense, the wall of container has contribution to the static pressure of liquid continuum. In fact, the local rotation will produce an isotropic compressive stress (which contributes the main body of static pressure) and surface contraction stress which is called as surface tension). In the theory, these results are obtained naturally, while in conventional mechanics theory, they are introduced externally. So, I have enough reasons to claim that the Chen rational mechanics theory frame is much powerful than others. Furthermore, the conventional static liquid definition fails to define its internal velocity field (which determines the local rotation in Chen theory). Hence, the static liquid definition given in this paper is much better.

Distinguishing from conventional static liquid continuum definition (gauge invariant), the static liquid is defined by motion within unit time. Hence, introducing the thermal parameter $\sigma_\Theta$ the static liquid under incremental temperature environment is described by $F_j^i = R_j^i(\Theta_0 + \sigma_\Theta)$ in statistical sense. Similarly, the static solid is expressed as $F_j^i = R_j^i(\sigma_\Theta)$. Though both have invariant gauge tensor, they are at different physical phase. Theoretically, there is no sharp boundary between both. Hence, the solid-liquid transition is formulated under the physical requirement of stress continuity. As the local rotation has no contribution to the gauge tensor, the macro external observable classical strain (hence, stress) variation is caused by local rotation. This feature is expressed as: zero intrinsic stretching (intrinsic strain) and non-zero classical strain. In physical sense, for solid-liquid transition, the material element is invariant (on gauge field sense). However, although the physical stress is continuous, the classical strain is not continuous. By the classical strain definition (strain rate, in most textbooks), the velocity field (displacement field in unit-time) is not continuous for solid-liquid transition.

For static gas, it is well known that to keep its fixed external configuration, a closed container must be used. Its internal expansive pressure is balanced by the container closed walls. For free state gas, the deformation is: $F_j^i = \frac{1}{\cos\theta}\tilde{R}_j^i$. (To keep the fixed macro external configuration, the container must supply the compressive stress to produce a deformation (in container scale): $F_j^i = (\cos\theta)\delta_j^i$).

Hence, the static gas is different from static liquid in the local rotation modes: for liquid, the gauge is invariant; for gas, the gauge is amplified. Hence, for the solid-gas transition, the solid must have a volume explosion to broken into material pieces.

However, for liquid-gas transition, the flow-ability of liquid may not require the liquid has such an volume explosion as the linkage between material elements of liquid as a continuum is weak (comparing with solid continuum). In liquid-gas transition, the stress field is continuous. However, the classical strain is not continuous. Hence, very complicated flow patterns are possibly be formed. Unfortunately, they are beyond the coverage of this paper as the dynamic process will be concerned.

Although the transition dynamic process is omitted in this research, the phase transition condition is well formulated. They give out a rational interpretation about the phase diagram of conventional materials. As a limit case, the super-fluid transition conditions are formulated, also.

Therefore, it can conclude that: (1) a geometrical field representation of solid, liquid, and gas as a continuum id established in rational mechanics frame; (2) a geometrical deformation theory of solid, liquid, and gas phases transition are formulated; (3) the conventional constitutive equations of solid, liquid, and gas are unified into a constitutive equation which express the stress by intrinsic strain.

As an example, the rock bursting is formulated by the phase formulation in this paper through formulating a suitable multi-phase model.

Theoretically, it is hoped that an unified geometrical field theory which represents the solid, liquid, and gas in the same motion concept will help to solve the multiphase continuum mechanics problem. However, this job is extremely difficult. This difficulty is apparent in this research, as the temperature is not given in a rational formulation although it is the representation of internal motion of matter. However, "step by step" is the only way to attend to the idea target.

**Reference:**
[1] Chen Zhida, Rational Mechanics, Xuzhuo: China University of Mining & Technology Press., 1987 (In Chinese)
[2] Truesdell, C., The Mechanical Foundation of Elasticity and Fluid Mechanics. New York: Gordon and Breach

Science Pub. Inc., 1966

[3] Xiao Jianhua. Evolution of Continuum from Elastic Deformation to Flow. E-print, arXiv: physics/0511170(physics.class-ph), 2005, 1-25

[4] Xiao Jianhua. Chen Rational Mechanics I: Introduction, Sciencepaper Online, http://www.paper.edu.cn, 2007-01-30

[5] Xiao Jianhua. Chen Rational Mechanics II: Geometrical Equations, Sciencepaper Online, http://www.paper.edu.cn, 2007-01-31

[6] Xiao Jianhua. Chen Rational Mechanics III: Deformation decomposition and strain definition, Sciencepaper Online, http://www.paper.edu.cn, 2007-02-06

[7] Xiao Jianhua. Chen Rational Mechanics IV: Constitutive Equations, Sciencepaper Online, http://www.paper.edu.cn, 2007-02-06

[8] Xiao Jianhua. Chen Rational Mechanics V: Motion Equations, Sciencepaper Online, http://www.paper.edu.cn, 2007-02-07

[9] Xiao Jianhua. Chen Rational Mechanics VI: Micro geometry and Macro Geometry, Sciencepaper Online, http://www.paper.edu.cn, 2007-03-09

[10] Xiao Jianhua. Chen Rational Mechanics VII: Viscoelastisity, Sciencepaper Online, http://www.paper.edu.cn, 2007-03-09

[11] Xiao Jianhua. Chen Rational Mechanics VIII: Fatigue and Cracking, Sciencepaper Online, http://www.paper.edu.cn, ,2007-03-19

[12] Xiao Jianhua. Chen Rational Mechanics IX: Dynamic Instability and Cracking Deformation, Sciencepaper Online, http://www.paper.edu.cn, 2007-03-12

[13] Xiao Jianhua. Chen Rational Mechanics X: Description of Path-dependence for one dimension deformation, Sciencepaper Online, http://www.paper.edu.cn, 2007-03-19

[14] Xiao Jianhua. Chen Rational Mechanics XI: Multi-scale and Locality of Periodic Structure of Solid, Sciencepaper Online, http://www.paper.edu.cn,, 2007-04-04

[15] Xiao Jianhua. Chen Rational Mechanics XII: Determine the elastic parameters from stress-strain experiment curves, Sciencepaper Online, http://www.paper.edu.cn, 2007-05-04

[16] Xiao Jianhua, Intrinsic Knots Produced by Large Deformation in 3-Space I: Curvatures, Sciencepaper Online, http://www.paper.edu.cn, 2007-10-17

[17] Xiao Jianhua, Intrinsic Knots Produced by Large Deformation in 3-Space II: Multi-scale, Sciencepaper Online, http://www.paper.edu.cn, 2008-1-10

[18] Xiao Jianhua. Decomposition of Displacement Gradient and Strain Definition. Advances in Rheology, J. Central South University of Technology, Vol.14 (Suppl.1), 401-404, 2007

[19] Xiao Jianhua, Determining Loading Field based on Required Deformation for Isotropic Hardening Materials, J. Shanghai Jiaotong Uni. (Science), E12(6):805-812, 2007

[20] Xiao Jianhua, Bubble dynamics equations for Newton fluid, ISND2007, *Journal of Physics: Conference Series* **96,** 2008, 012134, doi:10.1088/1742-6596/96/1/012134, 2008

[21] Xiao Jianhua. Motion Equation of Vorticity for Newton Fluid. E-print, arXiv: physics/0512051(physics.flu-dyn), 2005, 1-7

[22] Xiao Jianhua. Intrinsic Structure of Turbulent Flow in Newton Fluid. E-print, arXiv: physics/0601006(physics.flu-dyn), 2006, 1-16

[23] Lodge, A.S., Body tensor fields in continuum mechanics, Academic Press, New York, 1974